\documentstyle[aps,psfig,multicol]{revtex}
\begin{document}
\draft
\title{Electromechanical noise in a diffusive conductor}
\author{A. V. Shytov$^{1}$, L. S. Levitov$^{2}$, and C. W. J. Beenakker$^{3}$}
\address{$^{1}$Institute for Theoretical Physics, University of California,
Santa Barbara, CA 93106--4030}
\address{$^{2}$Department of Physics, Center for Materials Science \&
Engineering,\\
Massachusetts Institute of Technology, Cambridge, MA 02139}
\address{$^{3}$Instituut-Lorentz, Universiteit Leiden, P.O. Box 9506, 2300 RA
Leiden, The Netherlands}
\date{29 October 2001}
\maketitle
\begin{abstract}
Electrons moving in a conductor can transfer momentum to the lattice via
collisions with impurities and boundaries, giving rise to a fluctuating
mechanical stress tensor. The root-mean-squared momentum transfer per
scattering event in a disordered metal (of dimension $L$ greater than the mean
free path $l$ and screening length $\xi$) is found to be reduced below the
Fermi momentum by a factor of order $l/L$ for shear fluctuations and
$(\xi/L)^{2}$ for pressure fluctuations. The excitation of an elastic bending
mode by the shear fluctuations is estimated to fall within current experimental
sensitivity for a nanomechanical oscillator.
\end{abstract}
\pacs{PACS numbers: 85.85.+j, 73.23.-b, 73.50.Td, 77.65.-j}
\begin{multicols}{2}
\makeatletter
\newbox\brackd \newbox\bracku
\setbox\brackd=\hbox {\vrule height 0pt depth
  3pt width 1pt}\ht\brackd=0pt\dp\brackd=1pt
\setbox\bracku=\hbox {\vrule height 3pt depth
  0pt width 1pt}\ht\bracku=1pt\dp\bracku=0pt
\def\downbrackfill{$\m@th\copy\brackd\leaders\vrule\hfill\copy\brackd$}
\def\overbrack#1{\mathop{\vbox{\ialign{##\crcr\noalign{\kern3\p@}
      \downbrackfill\crcr\noalign{\kern3\p@\nointerlineskip}
      $\hfil\displaystyle{\kern1pt#1\kern1pt}\hfil$\crcr}}}\limits}
\makeatother
Impressive advances in the fabrication of miniature mechanical oscillators
provide new opportunities for research in mesoscopic physics
\cite{Sid95,Rou01}. The coupling of electrical and mechanical degrees of
freedom is of particular interest. We mention the observation of thermal
vibration \cite{Tre96} and acoustoelectric effects \cite{Reu00} in carbon
na\-no\-tubes, the coupling of the centre-of-mass motion of ${\rm C}_{60}$
molecules and single-electron hopping \cite{Par00}, and also theoretical work
\cite{Boc88} on the coupling between a tunneling electrical current and a
localized phonon mode.

The present paper was motivated by a question posed to us by M. Roukes:
Electrons in a metal collide with impurities and thereby exert a fluctuating
force on the lattice. In equilibrium this electromechanical force can not be
distinguished from other sources of thermal noise. Might it be measurable out
of equilibrium by driving a current through a nanoscale oscillator? To address
this question one has to consider a delicate balance of forces.

\begin{figure}
\hspace*{\fill}
\psfig{figure=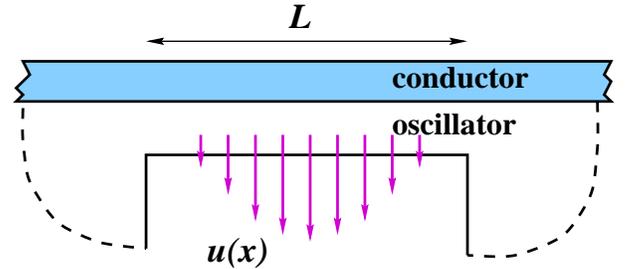,width=8cm}
\bigskip\\
\hspace*{\fill}
\caption[]{
Sketch of an elastic beam clamped at both ends to a contact and covered by a
metal layer. A current flowing through the metal excites a bending mode $u(x)$
of the beam.
\label{figoscillator}
}
\end{figure}

We will provide both a general theory and a specific application to the
electromechanical excitation of a bending mode in the geometry of Fig.\
\ref{figoscillator}: A thin elastic beam connecting two massive Ohmic contacts.
The beam could be a conductor or an insulator covered with a metal ({\em e.g.\
} a metallized suspended silicon beam \cite{Cle96}). We calculate the excess
noise in the bending mode that arises in the presence of a dc voltage $V$ and
conclude that it should be observable on the background of the thermal noise.

Let us first discuss the order of magnitude. The noise at low temperatures is
due to the ${\cal N}eV/E_{\rm F}$ ``noisy'' electrons within a range $eV$ of
the Fermi energy $E_{\rm F}$ (with $\cal N$ the total electron number in the
metal). Each electron transfers to the lattice a typical momentum $\Delta
p\simeq p_{\rm F}$ in a scattering time $\tau$. The mean squared momentum
transfer in a time $t$ for {\em uncorrelated\/} increments $\Delta p$ would be
\begin{equation}
({\cal N}eV/E_{\rm F})(\Delta p)^{2}(t/\tau)\simeq{\cal
N}m^{\ast}eV(t/\tau)\equiv {\cal P}_{\rm max}t,\label{Pmaxdef}
\end{equation}
with $m^{\ast}$ the electron effective mass.

We find that ${\cal P}_{\rm max}$ overestimates the fluctuations in the
transverse force. The actual noise power is of order ${\cal
P}\simeq(l/L)^{2}{\cal P}_{\rm max}$, with $L$ the length of the beam and $l$
the mean free path in the metal. The reduction is due to the fact that
subsequent momentum transfers are strongly correlated, basically because an
electron being scattered back and forth alternatingly transfers positive and
negative momentum to the lattice. The factor $(l/L)^{2}$ reduces the noise
substantially, but we estimate that it should be observable in an oscillator
with a $10^{-16}\,{\rm N}/\sqrt{\rm Hz}$ sensitivity \cite{Rou01,Cle96}.

We combine two independently developed theoretical frameworks: The dynamic
theory of elasticity \cite{Gur60,Kon71} and the kinetic theory of fluctuations
\cite{Kog69}. We start from the Boltzmann-Langevin equation of Kogan and
Shulman. This is a kinetic equation with a fluctuating source $\delta J({\bf
r},{\bf p},t)$, that describes fluctuations in time of the distribution
function $n({\bf r},{\bf p},t)$,
\begin{equation}
(\partial_{t}+{\bf v}\cdot\nabla_{\bf r}+e{\bf E}\cdot\nabla_{\bf p}+{\cal
S})n=\delta J.\label{Langevin}
\end{equation}
Here ${\bf p}=m^{\ast}{\bf v}$ is the quasimomentum and ${\bf E}({\bf r},t)$ is
the electric field. The collision integral ${\cal S}$ for elastic scattering on
impurities (with rate $W$) is given by
\begin{equation}
{\cal S}n({\bf p})=\langle W(\hat{\bf p}\cdot\hat{\bf p}')[n({\bf p})-n({\bf
p}')]\rangle_{\hat{\bf p}'}.\label{calSdef}
\end{equation}
The angular brackets indicate an average over the direction $\hat{\bf p}'$ of
the momentum ${\bf p}'$, with $|{\bf p}'|=|{\bf p}|$.

The noise source $\delta J$ has zero time average and variance
\begin{eqnarray}
&&\overline{\delta J({\bf r},{\bf p},t)\delta J({\bf r'},{\bf
p}',t')}=\delta({\bf r}-{\bf
r}')\delta(t-t')\delta(\varepsilon-\varepsilon')\nu^{-1}\nonumber\\
&&\hspace{1cm}\mbox{}\times\bigl[4\pi\delta(\hat{\bf p}-\hat{\bf p}')\langle
W(\hat{\bf p}\cdot\hat{\bf
p}'')(\bar{n}+\bar{n}''-2\bar{n}\bar{n}'')\rangle_{\hat{\bf p}''}\nonumber\\
&&\hspace{1cm}\mbox{}-W(\hat{\bf p}\cdot\hat{\bf
p}')(\bar{n}+\bar{n}'-2\bar{n}\bar{n}')\bigr].
\label{vardeltaJ}
\end{eqnarray}
Here $\nu(\varepsilon)$ is the density of states (at energy
$\varepsilon=p^{2}/2m^{\ast}$) and $\bar{n}$ is the time-averaged distribution.
We have set Planck's constant $h\equiv 1$, so that $n$ is dimensionless, and
abbreviated $\bar{n}'=\bar{n}({\bf r},{\bf p}',t)$, $\bar{n}''=\bar{n}({\bf
r},{\bf p}'',t)$.

The force density ${\bf f}({\bf r},t)$ exerted by the electrons on the lattice
is the divergence of a symmetric tensor $\mathbf\Pi$ that can be decomposed
into an isotropic pressure $\Pi_{0}$ and a traceless shear tensor
$\mathbf\Sigma$:
\begin{equation}
f_{\alpha}=-\nabla_{\beta}\Pi_{\alpha\beta}, \;\;\Pi_{\alpha\beta}=
\Pi_{0}\delta_{\alpha\beta}+\Sigma_{\alpha\beta}.\label{falphaPidef}
\end{equation}
In the approximation of a deformation-independent effective mass one has
\cite{Kon71,Fik78}
\begin{equation}
\Pi_{\alpha\beta}=m^{\ast}\int d{\bf p}\,v_{\alpha}v_{\beta}n.\label{Fdef}
\end{equation}

The time-averaged force (\ref{falphaPidef}) vanishes, since it contains a
derivative of the spatially uniform time-averaged distribution $\bar{n}$.
Physically, this results from a cancellation between the electrical current
drag on impurities (the socalled ``wind force'') and the electric field force
exerted on the ions \cite{Kag77}. Moreover, since $\bf f$ is a total derivative
the net fluctuating force vanishes as well at low frequencies (ignoring
boundary contributions). Although the center of mass does not move, there are
fluctuating compression modes (driven by $\Pi_{0}$) as well as torsion and
bending modes (driven by $\mathbf\Sigma$). The driving force ${\cal F}(t)$ for
each of these modes is obtained by weighing ${\bf f}({\bf r},t)$ with a
sensitivity function ${\bf g}({\bf r})$ proportional to the displacement field
of the mode,
\begin{equation}
{\cal F}=\int d{\bf r}\,{\bf f}\cdot{\bf g}=\int d{\bf
r}\,\left(\Pi_{0}\nabla\cdot{\bf g}+ \Sigma_{\alpha\beta}\frac{\partial
g_{\beta}}{\partial r_{\alpha}}\right).\label{calFurelation}
\end{equation}

The two contributions $\Pi_{0}$ and $\mathbf\Sigma$ can be separated by
expanding $n({\bf r},{\bf p},t)$ in spherical harmonics ${\bf n}^{(q)}({\bf
r},\varepsilon,t)$ with respect to the direction $\hat{\bf p}$ of the momentum.
It is convenient to write the spherical harmonics in cartesian (rather than
spherical) coordinates,
\begin{eqnarray}
n&=&\sum_{q=0}^{\infty}\overbrack{\hat{p}_{\alpha_{1}}
\cdots\hat{p}_{\alpha_{q}}} n^{(q)}_{\alpha_{1}\cdots\alpha_{q}}\nonumber\\
&=&n^{(0)}+\hat{p}_{\alpha}n^{(1)}_{\alpha}+(\hat{p}_{\alpha}\hat{p}_{\beta}-
\case{1}{3}\delta_{\alpha\beta})n^{(2)}_{\alpha\beta}+\cdots\label{nexpansion}
\end{eqnarray}
Here $\overbrack{\hat{\bf p}^{q}}$ is the traceless part of the symmetric
tensor $\hat{p}_{\alpha_{1}}\cdots\hat{p}_{\alpha_{q}}$. These tensors form an
orthonormal set \cite{Hes80},
\begin{equation}
\langle\overbrack{\hat{\bf p}^{n}}\overbrack{\hat{\bf p}^{m}}\rangle_{\hat{\bf
p}}=\delta_{nm}\frac{m!}{(2m+1)!!}{\mathbf\Delta}^{(m)}.\label{orthogonal}
\end{equation}
The tensor ${\mathbf\Delta}^{(m)}$ projects onto the traceless symmetric part
of a tensor of rank $m$. We will need
$\Delta^{(1)}_{\alpha\beta}=\delta_{\alpha\beta}$ and
\begin{equation}
\Delta^{(2)}_{\alpha\beta\alpha'\beta'}=
\case{1}{2}\delta_{\alpha\alpha'}\delta_{\beta\beta'}+
\case{1}{2}\delta_{\alpha\beta'}\delta_{\beta\alpha'}-
\case{1}{3}\delta_{\alpha\beta}\delta_{\alpha'\beta'}. \label{Deltadef}
\end{equation}

In view of the orthogonality of different spherical harmonics, one has
\begin{equation}
\Pi_{0}=\case{1}{3}\int d\varepsilon\,2\varepsilon\nu
n^{(0)},\;\;\Sigma_{\alpha\beta}=\case{2}{15}\int d\varepsilon\,2\varepsilon\nu
n^{(2)}_{\alpha\beta}.\label{PiSigman}
\end{equation}
The two harmonics $n^{(0)}$ and ${\bf n}^{(2)}$ have to be found from the
kinetic equation (\ref{Langevin}). We first consider the harmonic ${\bf
n}^{(2)}$ that determines the shear tensor ${\mathbf\Sigma}$, and then discuss
the harmonic $n^{(0)}$ and resulting pressure $\Pi_{0}$.

To obtain an equation for ${\bf n}^{(2)}$ we multiply both sides of Eq.\
(\ref{Langevin}) by $\overbrack{\hat{\bf p}\hat{\bf p}}$ and perform an angular
average. Employing the diffusion approximation on length and time scales larger
than $l$ and $\tau$, we neglect the derivatives with respect to $t$ and $\bf r$
in Eq.\ (\ref{Langevin}). Also, in the linear response approximation, we
neglect the derivative with respect to $\bf p$, since it gives a term bilinear
in $\bf E$ and $\bf j$. What remains is a local relation between ${\bf
n}^{(2)}$ and the second harmonic ${\bf J}^{(2)}$ of the fluctuating source,
\begin{equation}
{\bf n}^{(2)}=\tau_{2}\delta{\bf J}^{(2)}.\label{n2J2}
\end{equation}
The momentum transport time $\tau_{2}$ is defined by
\begin{equation}
1/\tau_{2}=\case{3}{4}\int_{-1}^{1}d\xi\,W(\xi)(1-\xi^{2}).\label{tau2def}
\end{equation}
For anisotropic scattering the time $\tau_{2}$ is larger than the charge
transport time $\tau$, defined by $1/\tau=\frac{1}{2}\int_{-1}^{1}d\xi
\,W(\xi)(1-\xi)$. (For isotropic scattering $\tau_{2}=\tau=1/W$.)

The correlator of $\delta{\bf J}^{(2)}$ follows in the same way from Eq.\
(\ref{vardeltaJ}). In the diffusion approximation we replace $\bar{n}$,
$\bar{n}'$, $\bar{n}''$ by $\bar{n}^{(0)}$. Using Eq.\ (\ref{orthogonal}) we
arrive at
\begin{eqnarray}
&&\overline{\delta{\bf J}_{\alpha\beta}^{(2)}({\bf r},\varepsilon,t)\delta {\bf
J}_{\alpha'\beta'}^{(2)}({\bf
r}',\varepsilon',t')}=\frac{15}{\nu\tau_{2}}
{\mathbf\Delta}_{\alpha\beta\alpha'\beta'}^{(2)} \delta(t-t')\nonumber\\
&&\mbox{}\times\delta({\bf r}-{\bf
r}')\delta(\varepsilon-\varepsilon')\bar{n}^{(0)}({\bf
r},\varepsilon)[1-\bar{n}^{(0)}({\bf r},\varepsilon)].\label{vardeltaJ2}
\end{eqnarray}
Since $\bar{n}^{(0)}({\bf r},\varepsilon)$ differs from 0 or 1 only in a narrow
range near the Fermi level, we ignore the energy dependence of $\nu$ and
$\tau_{2}$ and evaluate them at $\varepsilon=E_{\rm F}$.

We quantify the transverse momentum noise through the correlator of the shear
tensor,
\begin{equation}
{\cal C}_{\alpha\beta\alpha'\beta'}^{(2)}({\bf r},{\bf
r'})=2\int_{-\infty}^{\infty}dt\,\overline{\Sigma_{\alpha\beta}({\bf
r},0)\Sigma_{\alpha'\beta'}({\bf r}',t)}.\label{calCdef}
\end{equation}
Combining Eqs.\ (\ref{PiSigman}), (\ref{n2J2}), and (\ref{vardeltaJ2}), we
obtain the result
\begin{eqnarray}
&&\mbox{\boldmath$\cal C$}^{(2)}({\bf r},{\bf r'})=\case{8}{15}(m^{\ast} v_{\rm
F}^{2})^{2}\tau_{2}\nu\delta({\bf r}-{\bf r}')K({\bf r}){\mathbf\Delta}^{(2)},
\label{C2result}\\
&&K({\bf r})=\int d\varepsilon\,\bar{n}^{(0)}({\bf
r},\varepsilon)[1-\bar{n}^{(0)}({\bf r},\varepsilon)]. \label{Trdef}
\end{eqnarray}
The kernel $K$ is given by \cite{Nag92} $K=(eV/L)x(1-x/L)$, where $V$ is the
voltage applied between the two contacts at $x=0,L$. The parabolic profile
$K(x)$ requires $k_{\rm B}T\ll eV$ and the absence of inelastic scattering.

We now turn to the pressure fluctuations. Instead of Eqs.\ (\ref{n2J2}) and
(\ref{vardeltaJ2}) we have the fluctuating drift-diffusion equation
\cite{Nag92}
\begin{eqnarray}
&&{\bf j}+D\nabla\rho-\sigma{\bf E}=e\tau\int d\varepsilon\,\nu v\delta{\bf
J}^{(1)}\equiv\delta{\bf I},\label{driftdiffusion}\\
&&\overline{\delta I_{\alpha}({\bf r},t)\delta I_{\beta}({\bf
r}',t')}=2\sigma\delta_{\alpha\beta}\delta({\bf r}-{\bf r}')\delta(t-t')K({\bf
r}),\label{variancedeltaI}
\end{eqnarray}
which relates the fluctuations in the charge density $\rho=e\int
d\varepsilon\,\nu n^{(0)}$ and the current density ${\bf j}=\frac{1}{3}e\int
d\varepsilon\,\nu v{\bf n}^{(1)}$. Once we know the charge density fluctuations
we can find the fluctuating pressure from
\begin{equation}
\Pi_{0}=(D/\mu)\delta\rho,\label{Pirhorelation}
\end{equation}
cf.\ Eq.\ (\ref{PiSigman}). We have introduced the diffusion constant
$D=\frac{1}{3}v_{\rm F}^{2}\tau$, the conductivity  $\sigma=e^{2}\nu D$, and
the mobility $\mu=e\tau/m^{\ast}$.

The correlator ${\cal C}^{(0)}$ of the pressure fluctuations is defined as in
Eq.\ (\ref{calCdef}), with $\mathbf\Sigma$ replaced by $\Pi_{0}$. To close the
problem we need the continuity equation, $\partial\rho/\partial
t+\nabla\cdot{\bf j}=0$, and the Poisson equation $\kappa\nabla\cdot{\bf
E}=\delta\rho$ (with $\kappa$ the dielectric constant). The time derivative of
$\rho$ in the continuity equation may be omitted in the low-frequency regime of
interest. The fluctuations in the electron density then obey
\begin{equation}
D\nabla^{2}\delta\rho-(\sigma/\kappa)\delta\rho=\nabla\cdot\delta{\bf
I}.\label{deltarhojrelation}
\end{equation}

The current fluctuations create a fluctuating charge dipole, that is screened
over a length $\xi=(\kappa D/\sigma)^{1/2}= (\kappa/e^{2}\nu)^{1/2}$. On length
scales $\gg\xi$ one may neglect the diffusion term in Eq.\
(\ref{deltarhojrelation}) and use the local relation \cite{Nag98}
$\delta\rho=-(\kappa/\sigma)\nabla\cdot\delta{\bf I}$.
Eqs.\ (\ref{variancedeltaI}) and (\ref{Pirhorelation}) then yield
\begin{equation}
{\cal C}^{(0)}({\bf r},{\bf
r}')=\frac{4\sigma\xi^{4}}{\mu^{2}}\frac{\partial}{\partial{\bf
r}}\cdot\frac{\partial}{\partial{\bf r}'}\delta({\bf r}-{\bf r}')K({\bf
r}).\label{C0result}
\end{equation}

The next step is to use the results (\ref{C2result}) and (\ref{C0result}) to
estimate the low-frequency noise power ${\cal
P}=2\int_{-\infty}^{\infty}dt\,\overline{{\cal F}(0){\cal F}(t)}$ of the
fluctuating force ${\cal F}(t)$ that drives a particular oscillator mode. To
that end the correlator (\ref{calCdef}) is integrated over ${\bf r}$ and ${\bf
r}'$, weighted by the sensitivity function of the mode as in Eq.\
(\ref{calFurelation}). For a bending mode we use Eq.\ (\ref{C2result}), which
gives the noise power
\begin{equation}
{\cal P}^{(2)}=\case{8}{15}(m^{\ast}v_{\rm F}^{2})^{2}\tau_{2}\nu\int d{\bf
r}\,K({\bf r}) {\mathbf\Delta}^{(2)}_{\alpha\beta\alpha'\beta'}\frac{\partial
g_{\beta}}{\partial r_{\alpha}}\frac{\partial g_{\beta'}}{\partial
r_{\alpha'}}.\label{Pbending}
\end{equation}
For a compression mode in a metal of size $\gg\xi$ we use Eq.\ (\ref{C0result})
and find
\begin{equation}
{\cal P}^{(0)}=\frac{4\sigma\xi^{4}}{\mu^{2}}\int d{\bf r}\,K({\bf
r})|\nabla\nabla\cdot{\bf g}|^{2}.\label{Pcompression}
\end{equation}

For an order of magnitude estimate, we take $K\simeq eV$, ${\bf g}\simeq 1$,
and  also estimate spatial derivatives by factors $1/L$ and the volume integral
by a factor $\cal V$. For simplicity we assume isotropic impurity scattering,
so that $\tau_{2}=\tau$. Then the noise power due to fluctuations in the shear
tensor is of order ${\cal P}^{(2)}\simeq (m^{\ast}v_{\rm F}^{2})^{2}\tau\nu
eV{\cal V}L^{-2}$ and the noise power due to pressure fluctuations is of order
${\cal P}^{(0)}\simeq \sigma\xi^{4}\mu^{-2}eV{\cal V}L^{-4}$. It is instructive
to write these two estimates in the same form, using the identities
$\sigma/e\mu=\frac{1}{3}m^{\ast}v_{\rm F}^{2}\nu={\cal N}/{\cal V}\equiv n_{\rm
e}$, with $n_{\rm e}$ the electron density. One finds
\begin{equation}
{\cal P}^{(0)}\simeq(\xi/L)^{4}{\cal P}_{\rm max},\;\;{\cal
P}^{(2)}\simeq(l/L)^{2}{\cal P}_{\rm max},\label{Pestimates}
\end{equation}
with ${\cal P}_{\rm max}={\cal N}m^{\ast}eV/\tau$ the noise power for
independent momentum transfers mentioned in the introduction.

The experimental observation of the shear tensor fluctuations looks more
promising than of the pressure fluctuations, firstly because $\xi$ is typically
$\ll (lL)^{1/2}$ so that ${\cal P}^{(0)}\ll{\cal P}^{(2)}$, and secondly
because a typical oscillator operates in a bending or torsion mode rather than
in a compression mode. For that reason we will now limit the more quantitative
calculation to ${\cal P}^{(2)}$. We consider a bending mode
$u(x)\cos\omega_{0}t$ in the geometry of Fig.\ \ref{figoscillator}. The
sensitivity function $g(x)=u(x)/u(x_{0})$ equals the displacement (in the
$y$-direction) normalized by the value at a reference point $x_{0}$. We choose
$x_{0}=L/2$, so that ${\cal F}$ is equivalent to a point force at the beam's
center. Eq.\ (\ref{Pbending}) now takes the form
\begin{equation}
{\cal P}=\case{4}{5}n_{\rm e}p_{\rm F}(l{\cal A}/L)\int_{0}^{L}
\frac{dx}{L}\,K(x)[Lg'(x)]^{2},\label{Pbending2}
\end{equation}
with ${\cal A}={\cal V}/L$ the cross-sectional area of the metal layer.

The wave equation for transverse waves is biharmonic, $d^{4}u/dx^{4}=k^{4}u$.
The solution for doubly clamped boundary conditions is \cite{LL}
\begin{eqnarray}
u(x)&=&(\sin kL-\sinh kL)(\cos kx-\cosh kx)\nonumber\\
&&\mbox{}-(\cos kL-\cosh kL)(\sin kx-\sinh kx),\label{uxbeam}
\end{eqnarray}
with the resonance condition $\cos kL\cosh kL=1$. We use the lowest resonance
at $kL=4.73$. Substituting $K=(eV/L)x(1-x/L)$ and integrating we obtain the
excess noise ${\cal P}=\case{4}{5}n_{\rm e}p_{\rm F}(l{\cal A}/L)\times
0.83\,eV$. If we insert values typical for a metal, $n_{\rm e}=10^{29}\,{\rm
m}^{-3}$, $p_{\rm F}=10^{-24}\,{\rm Ns}$, $l=100\,{\rm nm}$, and choose typical
dimensions ${\cal A}/L=10\,{\rm nm}$, then the force spectral density at
$V=1\,{\rm mV}$ is ${\cal P}=10^{-32}\,{\rm N}^{2}/{\rm Hz}$. This is well
above the thermal noise power at low temperatures (of order $10^{-34}\,{\rm
N}^{2}/{\rm Hz}$ at $T=1\,{\rm K}$ \cite{Rou01}).

It is instructive to apply the result (\ref{Pbending2}) to a system in thermal
equilibrium, when $K(x)=k_{\rm B}T$ for all $x$. In this case an independent
estimate of the noise ${\cal P}_{0}$ is provided by the fluctuation-dissipation
theorem: ${\cal P}_{0}=4k_{\rm B}TM\omega_{0}/Q_{0}$, with $M$ the active mass
of the oscillator and $1/Q_{0}$ the electromechanical contribution to the
inverse quality factor. Eq.\ (\ref{Pbending2}) gives
\begin{equation}
\frac{1}{Q_{0}}=\case{1}{5}(l/L)^{2}\frac{{\cal
N}m^{\ast}}{M\omega_{0}\tau}\int_{0}^{L}
\frac{dx}{L}\,[Lg'(x)]^{2}.\label{Q0bending}
\end{equation}
This electromechanical quality factor might be measurable in a superconducting
metal, as an increase in the overall quality factor when $T$ drops below the
critical temperature. One can also calculate $Q_{0}$ directly as an
``absorption of ultrasound'' by conduction electrons \cite{Kon71}, providing a
consistency check on our analysis. However, the nonequilibrium noise
(\ref{Pbending2}) with an $x$-dependent kernel $K(x)$ can not be obtained from
acoustic dissipation.

Before concluding we mention an altogether different mechanism for
electromechanical noise, which is the coupling of a fluctuating surface charge
$\delta q(t)$ on the metal to the electromagnetic environment. In the presence
of an electric field $E_{0}$ between the metal surface and the substrate ({\em
e.g.\ } due to a mismatch in work functions), the charge fluctuations will give
rise to a fluctuating transverse force with noise power
\[
{\cal P}_{\rm env}\simeq E_{0}^{2}\overline{\delta q^{2}}\simeq
E_{0}^{2}C^{2}\overline{\delta V^{2}}\simeq E_{0}^{2}C^{2}R\,{\rm max}\,(k_{\rm
B}T,eV).
\]
Here $C$ is the capacitance to the ground and $R=L/{\cal A}\sigma$ the
resistance of the metal. The ratio ${\cal P}_{\rm env}/{\cal P}^{(2)}\simeq
(E_0CL^2/el{\cal N})^2$ is quite small for typical parameter values. The reason
is that the environmental charge fluctuations are a surface effect, while the
whole bulk of the metal contributes to ${\cal P}^{(2)}$. Although the noise per
electron was found to be small in $l/L$, the total noise power ${\cal P}^{(2)}$
is big due to the large number ${\cal N}eV/E_{\rm F}$ of contributing
electrons.

In summary, we have addressed the fundamental question of the excitation of an
elastic mode in a disordered metal out of equilibrium, as a result of the
fluctuating momentum that an electrical current transfers to the lattice. The
effect is small but measurable. The characteristic linear dependence of the
electromechanical noise on the applied voltage should distinguish it from other
sources of noise. We believe that a measurement is not only feasible but worth
performing. Indeed, the nonequilibrium electric current noise has proven to be
a remarkably powerful tool in the study of transport properties \cite{Bla00},
precisely because it contains information that is not constrained by the
fluctuation-dissipation theorem. The noise considered here could play a similar
role for mechanical properties.

This work was motivated by discussions with M. Roukes during the Nanoscience
program at the Institute for Theoretical Physics in Santa Barbara. We thank M.
Kindermann, Yu.\ V. Nazarov, and B. Spivak for discussions and acknowledge
support by the National Science Foundation under Grant No.\ PHY99--07949 and
No.\ DMR98--08941 (MRSEC program), and by the Netherlands Science Foundation
NWO/FOM.

\end{multicols}
\end{document}